\documentclass[twocolumn,showpacs,preprintnumbers,amsmath,prl,amssymb]{revtex4}

\usepackage{graphicx}
\usepackage{dcolumn}
\usepackage{bm}

\begin{document}

\title{Noise Thermometry with Two Weakly Coupled Bose-Einstein Condensates}

\author{Rudolf Gati}
\author{B\"orge Hemmerling}
\author{Jonas F\"olling}
\author{Michael Albiez}
\author{Markus K. Oberthaler}
\affiliation{Kirchhoff-Institut f\"ur Physik, Universit\"at
Heidelberg, Im Neuenheimer Feld 227, D-69120 Heidelberg, Germany.
E-Mail: noisethermometry@matterwave.de}

\date{\today}

\begin{abstract}
Here we report on the experimental investigation of thermally
induced fluctuations of the relative phase between two
Bose-Einstein condensates which are coupled via tunneling. The
experimental control over the coupling strength and the
temperature of the thermal background allows for the quantitative
analysis of the phase fluctuations. Furthermore, we demonstrate
the application of these measurements for thermometry in a regime
where standard methods fail. With this we confirm that the heat
capacity of an ideal Bose gas deviates from that of a classical
gas as predicted by the third law of thermodynamics.
\end{abstract}


\maketitle

The generation of two independent matter-wave packets by splitting
a single Bose-Einstein condensate (BEC) is a well established
technique \cite{bib1,ref2,ref3} in the field of atom optics. New
phenomena arise if the two separated parts can still coherently
interact in analogy to Josephson junctions in condensed matter
physics \cite{ref4} and superfluid Helium Josephson weak links
\cite{ref41}. An advantage of the realization of weakly coupled
BEC in a double-well potential \cite{ref5} is the possibility to
observe the phase difference between the two macroscopic wave
functions directly. Our experimental investigation of this
relative phase reveals that it is not locked to zero but exhibits
fluctuations. Two fundamental types of fluctuations are discussed
in the literature, quantum fluctuations \cite{ref6} and thermally
induced fluctuations \cite{ref7}. In this letter we report on the
experimental investigation of thermal fluctuations of the relative
phase arising from the interaction of the BEC with its thermal
environment, which is always present.

The essential prerequisite for the investigation of these
thermally induced phase fluctuations is the ability to prepare a
BEC adiabatically in a symmetric double-well potential and to
adjust its temperature. In our experiments this is achieved by
splitting a single $^{87}$Rb BEC produced and trapped in an
optical dipole trap by slowly ramping up a barrier in the center.
The tunneling coupling is adjusted by the barrier height and its
strength can be deduced from numerical simulations of the BEC in
the trap using the model described in \cite{ref8}. The temperature
of the BEC is adjusted by holding the cloud in the trap, where due
to fluctuations of the trap parameters energy is transferred to
the atoms. Once the final temperature is reached a standing light
wave is ramped up generating a barrier in the center, leading to
an effective double-well trapping potential (upper part of
Fig.~1a).

\begin{figure}[h!]
\includegraphics[totalheight=5.5cm]{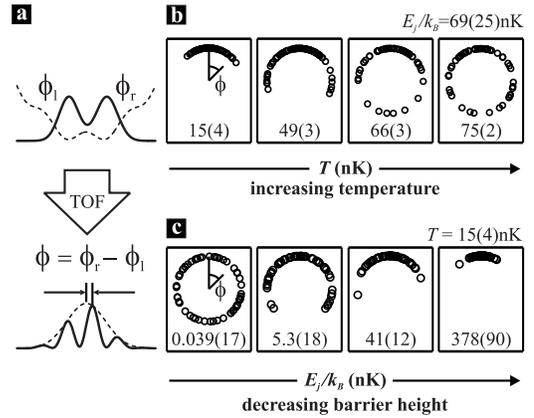}
\caption{\label{fig_1} Observation of thermal phase fluctuations.
The experimental steps are depicted in (\textbf{a}). A
Bose-Einstein condensate (solid line) is prepared in a double-well
potential (dashed line) by adiabatically ramping up the barrier.
The relative phase can be measured after a time-of-flight
expansion by analyzing the resulting double-slit interference
patterns (black line). (\textbf{b}) shows polar plots of the
relative phase obtained by repeating the experiment up to 60
times. The graphs show measurements for four different
temperatures \textit{T} at constant tunneling coupling energy
$E_j$, i.e. constant barrier height. The phase fluctuations
increase with increasing temperature. (\textbf{c}) shows polar
plots of the relative phase for a constant temperature at four
different tunneling coupling energies, i.e. different barrier
heights. Here the fluctuations are reduced with increasing
coherent tunneling coupling showing the stabilization.}
\end{figure}

When the potential is switched off the matter-wave packets start
to expand, overlap and form a double-slit interference pattern
which depends on their relative phase as indicated in the lower
part of Fig.~1a. Repeating the interference measurements reveals
that this relative phase is not constant but fluctuates around
zero. The general behavior of these phase fluctuations is
connected to two parameters: the temperature of the system
randomizing the phase and the tunneling coupling of the two
matter-wave packets stabilizing the phase. The results depicted in
Fig.~1b show that the phase fluctuations become more pronounced as
the temperature is increased since the fluctuations outweigh the
stabilizing effects. From this point of view it is expected - and
also experimentally observed (Fig.~1c) - that keeping the
temperature constant and increasing the tunneling coupling leads
to a reduction of the fluctuations. A measure for the fluctuations
is the coherence factor $\alpha=\langle\cos\phi\rangle$
\cite{ref7} which is directly connected to the visibility of the
ensemble averaged interference fringes.

\begin{figure}[h]
\includegraphics[totalheight=5.5cm]{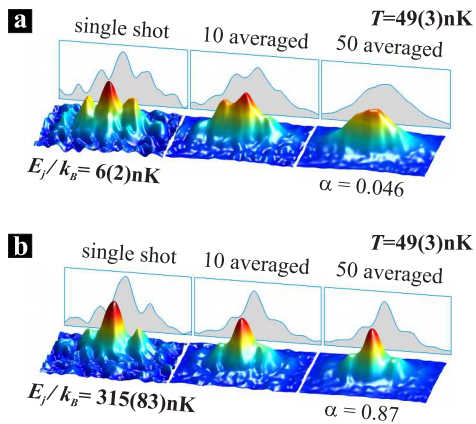}
\caption{\label{fig_2} Loss of the coherence of the bosonic
Josephson junction due to the coupling to a thermal environment.
(\textbf{a}) shows the transition from coherent single
realizations to incoherent ensemble averages. In the single
realization a clear interference signal is observed, where the
visibility is decreased due to the finite optical resolution of
the imaging setup. After averaging over 10 realizations the
visibility is reduced and after averaging over 50 realizations the
coherence is lost ($\alpha=0.046$). In (\textbf{b}) the coherent
evolution of the bosonic Josephson junction is depicted. The
results shown are obtained by repeating the experiment at the same
temperature as above but at a stronger tunneling coupling. Here
the averaging leads to only a small degradation of the visibility
($\alpha=0.87$) showing how coherent coupling can counteract
dephasing processes.}
\end{figure}

The coherence of the system can be visualized as shown in Fig.~2.
For every single realization below the critical temperature the
experiment reveals interference patterns with high visibility.
However, the visibility is reduced by averaging over many
realization and for high temperatures it disappears completely as
the mean fluctuations of the relative phase become comparable to
$\pi$. The loss of coherence due to thermal fluctuations is shown in
Fig.~2a corresponding to $\alpha=0.046$. At the same temperature the
coherence of the weakly coupled condensates can be maintained by
increasing the coupling. For a tunneling coupling energy larger than
the thermal energy the phase is locked to zero and the averaging
reduces the visibility only slightly as shown in Fig.~2b where the
coherence factor is given by $\alpha=0.87$. The dependence of the
coherence factor on the two parameters is the consequence of a
universal scaling law which can be explained by the classical model
discussed in the following.

The dynamics can be described in terms of a two mode approximation
by assuming weak coupling between the localized modes of the BEC.
This corresponds to treating the BEC in the double-well potential
as two separated matter-wave packets connected via tunneling
through the barrier. In the following we will use the acronym for
bosonic Josephson junction (BJJ) to describe this system. Within
the two mode approximation the dynamics of the BJJ can be
described by two conjugate variables, the atom number difference
between the matter-wave packet on the left (l) and on the right
(r) $\Delta n = (N_l - N_r)/2$ and their relative phase
$\phi=\phi_r-\phi_l$ \cite{ref8,ref9,ref10,ref11}. The Hamiltonian
governing the evolution of the two conjugate variables in the
limit of small $\Delta n$ is given by
\begin{equation}
H=\frac{E_c}{2}\Delta n^2 - E_j \cdot \cos \phi \quad ,
\end{equation}
where $E_c$ accounts for the atom-atom interaction in both
condensates and $E_j$ is the tunneling coupling energy resulting
from the spatial overlap of the wave functions. This Hamiltonian
also describes the classical motion of a particle with mass
$1/E_c$ and momentum $\Delta n$ at position $\phi$ in a periodic
potential. In our experiments with temperatures $T>10$nK the
quantum fluctuations \cite{ref7,ref12,ref13,ref14} are small
compared to the thermal fluctuations and therefore are neglected.
Their influence can be estimated in the limit of small $\phi$ in
which Eq.~(1) can be approximated by a harmonic oscillator with
the characteristic quantum mechanical energy splitting $\hbar
\omega_p = \sqrt{E_c \cdot E_j}$ where $\omega_p$ is the plasma
frequency, leading to the quantum mechanical fluctuations of both
variables: $\langle \Delta n^2\rangle\approx\sqrt{E_j/4E_c}$ and
$\langle \phi^2\rangle\approx\sqrt{E_c/4E_j}$.

The system variables $E_j$ and $E_c$ can be calculated from the
experimental parameters. The trapping frequencies of the
three-dimensional harmonic trap are $\omega_x=2\pi \cdot 90(2)$Hz
and $\omega_{y,z}=2\pi \cdot 100(2)$Hz. The periodic potential of
$V=V_0/2(1+\cos(2\pi/\lambda\cdot x))$ is realized by the
interference of two laser beams at a wavelength of $830$nm
crossing under an angle of $10^\circ$ resulting in a standing
light wave with periodicity of $\lambda=4.8(2)\mu$m and is ramped
up to a height of $V_0/h=$ 500Hz to 2500Hz. The number of atoms in
the BEC fraction is chosen to be 2500(500). After the preparation
of the BEC in the double-well trap the relative phase of the two
matter-wave packets is measured by analyzing the double-slit
interference patterns formed after time-of-flight of 5 and 6ms.
The visibility of these patterns is reduced due to the short
expansion time and the finite optical resolution of the imaging
system. Further details of the experimental setup can be found in
\cite{ref15}.

The relevant quantities can be calculated from these parameters
using the improved two mode model \cite{ref8}: $E_c/k_B$ is on the
order of 20pK and $E_j/k_B$ is between 30pK and 400nK \cite{ref16}
leading to $\hbar \omega_p / k_B$ being between 25pK and 3nK.
Thus, both necessary conditions for the classical limit are
fulfilled: $E_j \gg E_c$ leading to small quantum fluctuations of
$\phi$ and $E_c \gg E_j/N^2$ (where $N$ is the total number of
atoms in the BEC) leading to small quantum fluctuations of $\Delta
n/N$. Hence, our experiment can be discussed in the classical
framework where the thermally induced phase fluctuations are
closely analogous to the Brownian motion of a particle in a
sinusoidal potential.

\begin{figure}[h]
\includegraphics[totalheight=5.5cm]{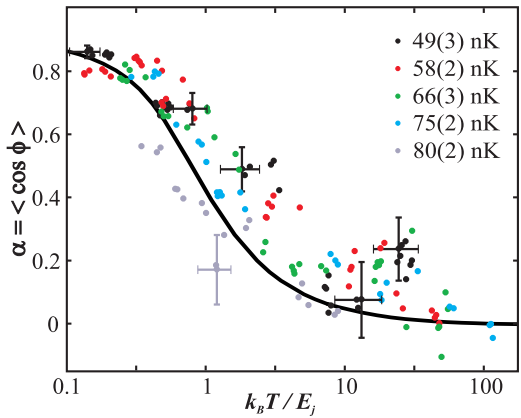}
\caption{\label{fig_3} Scaling behavior of the coherence factor of
a bosonic Josephson junction. Each point is obtained by averaging
over the cosine of the phases of at least 28 (in average 40)
measurements at the same experimental conditions. The coherence
factor $\alpha$ is plotted as a function of the scaling parameter
$k_BT/E_j$ which is varied over three orders of magnitude
($49\textrm{nK}<T<80\textrm{nK}$,
$0.6\textrm{nK}<E_j/k_B<300\textrm{nK}$). It shows good agreement
with the theoretical prediction of the classical model Eq.~(2)
indicated by the solid line where also the uncertainty arising
from the fitting error of the phase is taken into account. Typical
error bars are shown which result from statistical errors and
uncertainties of the experimental parameters (potential
parameters, atom numbers, temperature).}
\end{figure}

For a quantitative analysis in the thermodynamic limit at
$k_BT\gg\hbar\omega_p$ the coherence factor \cite{ref7} can be
calculated by a thermal average assuming a Boltzmann distribution
for the relative phases
\begin{equation}
\alpha=<\cos \phi>=\frac{\int_{-\pi}^{\pi}d\phi \cdot \cos \phi
\cdot \exp (E_j / k_B T \cdot \cos \phi)}{\int_{-\pi}^{\pi}d\phi
\cdot \exp (E_j / k_B T \cdot \cos \phi)} \quad .
\end{equation}
Eq. (2) points out that the relevant scaling parameter for thermal
fluctuations is the ratio between thermal energy $k_BT$ and
tunneling coupling energy $E_j$. Fig.~3 shows the experimentally
obtained coherence factors as a function of this scaling
parameter. Every data point represents on average 40 measurements.
In these experiments the temperature of the system is changed
between 49nK and 80nK by evaporatively cooling the sample to the
lowest temperature and subsequently increasing the temperature by
holding the atoms in the trap for different times. The temperature
of the sample is measured with the standard time-of-flight
expansion method. The tunneling coupling energy is varied between
0.6nK$\cdot k_B$ and 300nK$\cdot k_B$ by adjusting the height of
the potential barrier. $E_j$ is obtained from numerical
calculations using independently measured trap parameters and atom
numbers. It is important to note that the recently developed
improved two mode model \cite{ref8} is used for these calculations
because it leads to quantitative agreement between theoretical
predictions and experimental measurements of dynamical quantities
\cite{ref15}. The solid line corresponds to the theoretical
prediction of the classical model (Eq.~(2)) where all parameters
are determined independently. It also includes the fitting error
of the relative phase which arises from the finite optical
resolution and leads to a reduction of the coherence factor. As
shown in Fig.~3 the general behavior of the coherence is confirmed
over a three orders of magnitude variation of $k_BT/E_j$. These
measurements reveal that the BJJ has a higher degree of coherence
than expected. This deviation might possibly be explained by an
increase of the tunneling coupling resulting from the excitation
of transverse modes with higher energies which are neglected by
the two mode approximation.

Independent measurements have been performed for the lowest
temperatures ($T=15$nK) to test for thermal equilibration. The
measurements of $\alpha$ were compared for different $E_j$ for two
ramping schemes. The first scheme was ramping up the barrier in
1.3s and the second scheme was holding the atoms for 1s in the
trap and then ramping up the barrier within 0.3s. For
$E_j/k_B>1$nK both schemes lead within the experimental errors to
the same results. Thus for the fluctuation measurements the
ramping in 300ms is expected to be adiabatic with respect to the
response time of the BJJ given by the inverse plasma frequency and
thus ensures the thermal equilibrium.

In the following we present the application of the phase fluctuation
measurements for thermometry far below the critical temperature of
Bose-Einstein condensation ($T_c$). The temperature of the system
can be directly deduced from the variance of the phase if the
tunneling coupling is known. In order to apply the phase fluctuation
measurements for thermometry we introduce an empirical effective
tunneling coupling $E_{j}^{eff}$ to account for effects beyond the
classical approach. For the range of 25nK $< E_j/k_B <$ 90nK we
deduce from the results shown in Fig.~3 that $E_{j}^{eff}=1.33 \cdot
E_j$. The fundamental difference between this method and previous
suggestions using phase fluctuations of elongated Bose-Einstein
condensates for thermometry \cite{ref18} is that the BJJ is not
restricted to a quasi one dimensional situation but can be employed
for all geometries. Furthermore, this method can be applied for all
temperature ranges by tuning $E_c$ and $E_j$ such that thermal
effects dominate and quantum fluctuations are negligible.

\begin{figure}[h]
\includegraphics[totalheight=5.5cm]{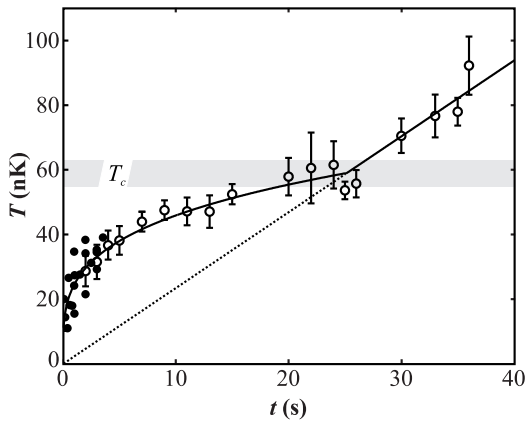}
\caption{\label{fig_4} Heating up of a Bose gas. The filled
circles correspond to measurements employing the phase fluctuation
method and the open circles to the results obtained with the
standard time-of-flight method. The grey shaded region shows the
critical temperature expected from the experimental parameters and
their uncertainties. The solid line is a fitting function assuming
a power law for the heat capacity $C \propto (T/T_c)^d$ of the
Bose gas below the critical temperature $T_c=59(4)$nK, a constant
heat capacity above the critical temperature and a temperature
independent transfer rate of energy. From this fit we deduce
$d=2.7(6)$ which is consistent with the theoretical prediction of
$d=3$ for an ideal Bose gas in a three-dimensional harmonic trap.
The dashed line represents the expected behavior of an ideal
classical gas for increasing temperature which makes the
difference arising from quantum statistics evident.}
\end{figure}

As a proof of applicability of this new type of thermometer we
observe how the temperature of a BEC in a harmonic trap increases
in time (see Fig.~4), which reveals clearly the effect of quantum
statistics below the critical temperature. In these experiments
the lowest temperatures ($T<T_c/3$) can only be measured with the
phase fluctuation method since the thermal fraction is too small
to be observed in time-of-flight measurements (less than 100 atoms
with about 2500 atoms in the BEC fraction). For longer heating
times the standard time-of-flight method can be applied and
confirms the consistency of the two approaches in the overlap
region. The solid line corresponds to a fitting function for the
temperature where we assume a mean critical temperature of
$T_c=59$nK (deduced from independent measurements), a temperature
independent transfer rate of energy per particle and a power law
for the temperature dependent heat capacity $C=(d+1) \cdot C_{th}
\cdot (T/T_c)^d$ where $C_{th}$ is the heat capacity of a
classical gas. The shown excellent agreement is obtained for a
heating rate of $2.3(2)$nK/s for a classical gas and $d=2.7(6)$.
Thus the expected exponent $d=3$ for an ideal Bose gas in a
three-dimensional harmonic trap \cite{ref19} is experimentally
confirmed. The expected increase of temperature of a classical gas
is indicated by the dotted line and shows clearly the difference
between the quantum and the classical behavior of ideal gases.

In summary, we have presented a quantitative analysis of thermally
induced phase fluctuations in a bosonic Josephson junction. Our
observations show that a universal scaling law describes the
behavior of the coherence and its control leads to new applications.
A method is presented for ultra-low temperature measurements, with
which we have confirmed that the heat capacity of a degenerate Bose
gas vanishes in the zero temperature limit as predicted by the third
law of thermodynamics \cite{ref20}.

\begin{acknowledgments}
We thank T. Bergeman very much for the numerical calculation of
the relevant parameters and the valuable theoretical support. We
would also like to thank B. Eiermann and T. B. Ottenstein for
discussions and M. Cristiani, Th. Anker and S. Hunsmann for their
contributions to the experimental setup. This work was funded by
Deutsche Forschungsgemeinschaft Schwerpunktsprogramm SPP1116 and
by Landesstiftung Baden-W\"urttemberg - Atomoptik. R. G.  thanks
the Landesgraduiertenf\"orderung Baden-W\"urttemberg for the
financial support.
\end{acknowledgments}


\begin{thebibliography}{}
\bibitem{ref16} The atom-atom interaction energy can be estimated by
$E_c/k_B \approx 51\rm{nK}/ N \cdot(1 - \exp
(\frac{-V_0/k_B-27\rm{nK}}{41\rm{nK}}))$ and the tunneling
coupling energy by $E_j/k_B \approx 196\rm{nK}\cdot N \cdot \exp
(\frac{-V_0/k_B-27\rm{nK}}{7.6\rm{nK}})$ for $N \approx 2500$ and
$24\rm{nK}<V_0/k_B<84\rm{nK}$ which is deduced from numerical
simulations using \cite{ref8}.
\end{thebibliography}
\end{document}